\documentclass[a4paper,preprint, superscriptaddress,prl,floatfix]{revtex4}
\usepackage{epsfig}
\usepackage{epstopdf}
\usepackage{graphicx}
\usepackage{amsfonts,amsmath,latexsym,amssymb}

\begin{document}

\newcommand{\vect}[1]{{\bf #1}}
\newcommand{\comment}[1]{\hfill {\tt {#1}}}

\title{
Inhomogeneous magnetic phases:  a LOFF-like phase in Sr$_3$Ru$_2$O$_7$}

\author{A. M. Berridge}
\author{A. G. Green}
\affiliation{School of Physics and Astronomy, University of St Andrews,
North Haugh, St Andrews KY16\ 9SS, UK}

\author{S. A. Grigera}
\affiliation{School of Physics and Astronomy, University of St Andrews,
North Haugh, St Andrews KY16\ 9SS, UK}

\affiliation{Instituto de F\'{\i}sica de L\'{\i}quidos y Sistemas
  Biol\'ogicos, UNLP, La Plata 1900, Argentina} 

\author{B. D. Simons}
\affiliation{Cavendish Laboratory, University of Cambridge, Madingley Road,
Cambridge, CB3\ 0HE, UK}

\date{\today}

\begin{abstract}
The phase diagram of Sr$_3$Ru$_2$O$_7$ contains a metamagnetic transition that bifurcates to enclose an anomalous phase with intriguing properties - a large resistivity with anisotropy that breaks the crystal-lattice symmetry. We propose that this is a magnetic analogue of the spatially inhomogeneous superconducting Fulde-Ferrel-Larkin-Ovchinnikov state. We show - through a Ginzburg-Landau expansion where the magnetisation transverse to the applied field can become spatially inhomogeneous -  that a Stoner model with electronic band dispersion can reproduce this phase diagram and transport behavior. 
\end{abstract}

\maketitle

Fulde and Ferrell~\cite{Fulde} and Larkin and Ovchinnikov~\cite{LarkinOvchinnikov} conjectured that the transition between superconducting and insulating behavior, driven by a magnetic field, could occur {\it via} an intermediate phase with spatially modulated superconducting order. 
This
proposal has since been extended to a wide range of settings, from ultracold atomic Fermi gases~\cite{cold_atoms} and exciton insulators~\cite{Balents} to quark matter and neutron stars~\cite{Loff_review}. However, experimental confirmation of these predictions is still controversial~\cite{LOFF_exp, LOFF_exp_2}.
In a similar spirit, intermediate phases between a Fermi liquid and Wigner crystal~\cite{Spivak} have been discussed.
We propose an inhomogeneous magnetic phase that can be considered a magnetic analogue of the LOFF phase. In this case, a change in homogeneous ferromagnetic order occurs {\it via} an intermediate phase with spatially modulated magnetization. This phase would generate clear experimental signatures. Furthermore, we argue on the basis of both new and previous experimental results that the anomalous phase behaviour observed in Sr$_3$Ru$_2$O$_7$~\cite{Grigera04,Green05,Borzi07,Grigera03a,Grigera01} can be explained in this way.

\begin{figure}[ht]
\includegraphics[height=2in]{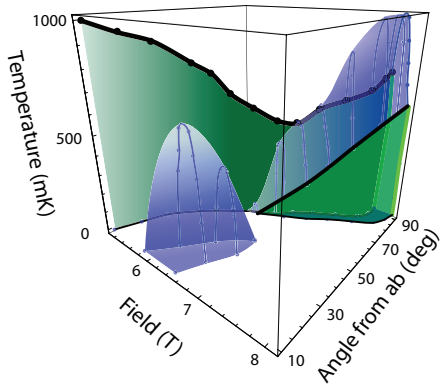}
\caption{
 The phase diagram of Sr$_3$Ru$_2$O$_7$ as inferred from in-plane transport properties. The green planes correspond to abrupt changes in resistivity as a function of field. Blue shading indicates regions where the in-plane resistivity is anomalously high, becomes highly anisotropic with respect to the in-plane component of the field~\cite{Borzi07}, and shows an anomalous temperature dependence.
The phase diagram obtained from magnetic susceptibility~\cite{Green05} shows the same first order transitions as indicated here in green, but lacks the roof.}
\label{fig:Experiment}
\end{figure} 

The bilayered ruthenate Sr$_3$Ru$_2$O$_7$ shows a sequence of metamagnetic transitions~\cite{Grigera03a}. Recent ARPES data has found evidence of van Hove singularities that may drive this metamagnetism~\cite{Felix}. Early studies focussed on a line of metamagnetic critical end-points that could be tuned to a quantum critical point by adjusting the magnetic field strength and orientation~\cite{Grigera01}. Subsequently, ultra-pure samples showed a bifurcation of this line upon approaching the putative quantum critical point~\cite{Grigera04,Green05} with a second line of critical end-points emerging from the zero-temperature plane (see Fig.\ref{fig:Experiment}). This bifurcation is accompanied by the appearance of a striking peak in resistivity~\cite{Grigera04} with curious, anisotropic dependence on the relative orientation of current, lattice and in-plane magnetic field~\cite{Borzi07}. When current flows in the crystallographic direction most parallel to the in-plane field, the resistivity peak rapidly decreases as the field is moved away from the c-axis. When it is nearly perpendicular to the in-plane field, the peak persists.  Further indications of a ``roof'' delineating the region of anomalous phase behavior with field along the c-axis~\cite{Grigera04} were provided by a kink in the longitudinal magnetization and a qualitative change in the temperature dependence of resistivity. Fig.~\ref{fig:Experiment} uses new resistivity data to extend this roof in angle. Similar features occur elsewhere in the phase diagram~\cite{Borzi07}, with further bifurcations apparent upon approaching the ab-plane. These show a smaller resistance anomaly, but have the same characteristic anisotropy.


Beginning with a heuristic discussion of the physics of the LOFF state and its magnetic analogue, in the following, we will describe how the the Wohlfarth-Rhodes~\cite{Wohlfarth} band picture of metamagnetism is extended to allow the possibility of spatially modulated magnetic phases. In order to deduce the effects upon the broader phase diagram, we turn to a Ginzburg-Landau expansion of the microscopic Hamiltonian. The key physics is revealed in an expansion along the line of metamagnetic critical end-points through a vanishing stiffness to spatial modulation of the transverse magnetization. This leads to a reconstruction of the phase diagram. Finally, we describe how our picture explains the behaviour of Sr$_3$Ru$_2$O$_7$---capturing both the experimental phase diagram and the properties of the anomalous phase.

A BCS superconductor is formed by binding electrons at the Fermi surface with opposite spin and momentum (${\bf k}, \uparrow$ and $-{\bf k},\downarrow$) to form Cooper pairs. A magnetic field imposes a Zeeman energy cost on the superconductor which is balanced against the condensation energy. When Zeeman energy dominates, the superconducting state is destroyed; Cooper pairs are broken allowing a spin polarization to develop. The transition from a superfluid to a normal phase can occur {\it via} an intermediate inhomogeneous condensate, the LOFF phase~\cite{Fulde, LarkinOvchinnikov}. By pairing electrons into a state with non-zero total momentum (${\bf k}+{\bf q}/2,\uparrow$ and $-{\bf k}+{\bf q}/2,\downarrow$), 
the reduction in condensation energy due to modulation is offset by a gain in Zeeman energy. The precise texture of the superconducting order
depends sensitively upon microscopic details~\cite{Loff_review}. 

A similar mechanism can apply to itinerant magnets: A spatially-modulated magnetic phase may intervene between the high- and low-magnetization states of a metamagnet. To form a ferromagnet, there must be an energetic gain in transferring an electron from a spin-down to a spin-up state of the same momentum. In a Stoner model, this is due to Coulomb exchange energy acquired at the expense of kinetic energy. Extending the Stoner model to include a band dispersion with peaks in the electronic density of states (DoS) leads to metamagnetism~\cite{Wohlfarth, Binz04}: As the Fermi surface of, say, majority carriers approaches its van Hove filling, the single-particle energy cost in changing its filling is reduced. This can lead to a step change in the magnetization at certain values of the external
field. 

\begin{figure}[tb]
\center{\includegraphics[width=2.75in]{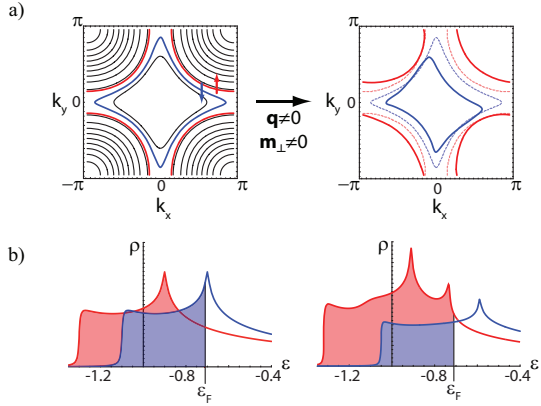}}
\caption{
a) Energy contours for a next-nearest neighbor tight-binding model with Fermi 
surfaces for minority and majority electrons shown in blue and red, 
respectively (left).
The minority and majority bands are distorted by a spiral modulation with 
non-zero transverse magnetization and wavevector (right).
b) DoS of minority and majority
states with a uniform magnetization (left) and with a spiral distortion 
(right). In the former, 
the Fermi surface lies just below a peak in the 
DoS and in the latter it lies between two split peaks.}
\label{fig:Cartoon2}
\end{figure}

Inhomogeneous magnetic states can be stabilized by peaks in the DoS in a similar way to spin density waves~\cite{Gil, Rice}. The simplest inhomogeneous phase formed from a ferromagnet is a spin spiral~\cite{foot}. A spiral of the right wavevector distorts the Fermi surface so that some regions are brought closer to their van Hove filling (see Fig.~\ref{fig:Cartoon2}). The reduction in single particle energy cost due to occupying states near to the peak in the DoS can outweigh the single-particle energy costs from elsewhere. This leads to peaks in the transverse magnetic susceptibility~\cite{Gil} and ultimately provides a mechanism by which a metamagnetic transition can split the transition between low and high magnetization occurring {\it via} a phase of inhomogeneous transverse magnetization. 

Such behavior can be shown explicitly for a Stoner model with
band dispersion: ${\hat {\cal H}}= \sum_{{\bf k},\sigma=\uparrow, \downarrow} 
\epsilon_{\bf k} \hat n_{{\bf k}, \sigma}- U \hat n_\uparrow \hat n_\downarrow 
- \mu_B H (\hat n_\uparrow- \hat n_\downarrow)$, where $\hat n_{{\bf k}, \sigma}$ is the number operator for electrons with momentum ${\bf k}$ and spin $\sigma$ and $\hat n_{\uparrow, \downarrow}$ is the total number operator for spin-up and spin-down electrons, respectively. $\epsilon_{\bf k}$ is the electronic dispersion--- we focus, without loss of generality, upon a two-dimensional tight-binding dispersion with next-nearest neighbor hopping. As noted above, this model displays metamagnetism~\cite{Wohlfarth, Binz04}.

Inhomogeneous phase formation leads to a reconstruction of the metamagnetic phase diagram that is best revealed through a Ginzburg-Landau expansion.
The thermodynamic properties of a metamagnetic 
system can be developed as a Landau expansion in magnetization density, 
${\bf M}$, as~\cite{Millis}
\begin{eqnarray}
\beta F_{\rm L}=r{\bf M}^2+u{\bf M}^4+v{\bf M}^6-{\bf h}\cdot{\bf M}\,,
\label{Hparent}
\end{eqnarray}
where ${\bf h}=h\hat{\bf e}_\parallel$ is the external magnetic field. 
The coefficients of this expansion for the Stoner model may be calculated in a standard manner following a perturbative expansion in interaction~\cite{Moriya}, and are specific functions of the external parameters; magnetic field, temperature and distance from van Hove filling. The point $r=u=h=0$ denotes the position of the parent tricritical point where the line of continuous Stoner transitions at zero field ($r=h=0$, $u>0$) bifurcates symmetrically into two lines of 
metamagnetic critical end-points parameterized by the conditions, 
$\partial_M {\cal F}_L=\partial_M^2 {\cal F}_L=\partial_M^3 {\cal F}_L
\stackrel{!}{=}0$. 

As we are interested in a reconstruction of the metamagnetic transition, it is convenient to shift our expansion from zero magnetization to an expansion about the magnetization along the line of metamagnetic critical end points. 
Setting ${\bf M}/ \bar M=\left(1+
\phi({\bf r})\right) \hat{\bf e}_\parallel+\boldsymbol\phi_\perp({\bf r})$, 
where $\bar{M}$ denotes the mean-field magnetization along the 
metamagnetic line~\cite{foot2},  and substituting into~(\ref{Hparent}) gives
\begin{eqnarray}
&&\frac{\beta F_{\rm L}}{h\bar{M}}=-H\phi+R\phi^2+\frac{5}{8}\phi^4
\nonumber\\
&&\qquad\qquad +\frac{1}{2} \left(1-\phi+\phi^2\right) \boldsymbol\phi_\perp^2
-\frac{1}{8}\boldsymbol\phi_\perp^4+\cdots
\label{H2}
\end{eqnarray}
$H$ and $R$ parameterize deviations from the metamagnetic critical 
end-point perpendicular and parallel to the first order line.  The 
dependence of the higher order coefficients on $H$ and $R$ can be neglected. 
Although we are interested in reconstructions of the metamagnetic transition that may lie outside the radius of convergence of the parent Landau theory~(\ref{Hparent}), it turns out that an explicit expansion for the Stoner model gives the same coefficients as (\ref{H2}) when constrained to lie along the parent line of critical end points~\cite{Andrew}. 

To allow for inhomogeneous
phase formation, we consider a minimal gradient expansion of the free energy:
\begin{eqnarray}
&&\beta F_{\rm GL}=\beta F_{\rm L}+\left(K_\perp + K_1 \phi +K_2 
\phi^2 +K_3 \boldsymbol\phi_\perp^2 \right)(\nabla \boldsymbol\phi_{\perp})^2
\nonumber\\
&&\qquad\qquad\qquad\qquad\qquad\qquad\qquad + L_\perp(\nabla^2 
\boldsymbol\phi_{\perp})^2\,,
\label{GLtheory}
\end{eqnarray}
where the parameters $K_1$, $K_2$, $K_3$ and $L_\perp$ are functions of the external parameters fixed by the microscopic theory. We have neglected gradient terms associated with $\phi$. While such terms can lead to
a spatial modulation, they do not lead to the phase reconstruction that 
we find. Gradient terms of fourth order and higher ought strictly to respect 
the lattice anisotropy~\cite{extra}. We consider the isotropic case for 
simplicity.

The key
ingredient introduced by explicit evaluation of the coefficients of the gradient expansion for the Stoner theory - that cannot be anticipated on purely symmetry grounds - is that $K_\perp$ changes sign 
along the line of metamagnetic critical end points~\cite{Kperp}. This indicates an instability to the formation of a spiral transverse magnetization. As this spiral order is established, the effective $\phi^4$ term changes sign leading to a tricritical point~\cite{Green05}.
\begin{figure}[hbt]
\centerline{\includegraphics[width=2.75in]{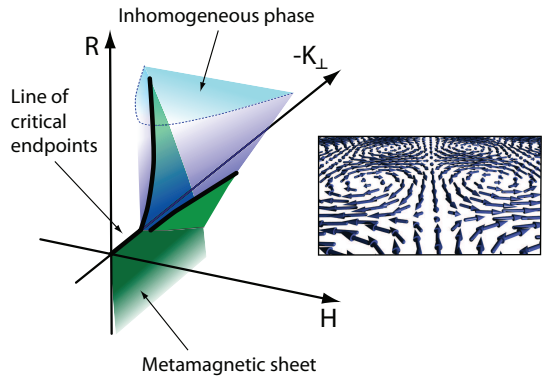}}
\caption{Phase diagram for the Ginzburg-Landau
  theory with possible spin texture.  Green sheets represent first-order transitions in $\phi$.  
  Blue sheets represent continuous transitions into the inhomogeneous phase.  The possible spin texture is constructed from four spin helices arranged in a square. The longitudinal magnetization has been supressed in this picture for emphasis. }
\label{fig:PhaseDiagram}
\end{figure}

The resulting phase diagram is shown in Fig.~\ref{fig:PhaseDiagram}. The metamagnetic 
sheet bifurcates at a dislocated (symmetry broken) tricritical 
point as shown in green~\cite{Green05}. The bifurcated wings embrace a 
region of inhomogeneous transverse magnetization in 
accord with our heuristic description. This region is further enclosed 
by a surface of continuous phase transitions, shown in blue,
 at which the transverse magnetization falls 
to zero. The longitudinal magnetization shows a kink on this 
surface--- a ghost of the transition in the transverse 
magnetization.

The inhomogeneous magnetic structure may consist of a superposition of several wavevectors. The sum of these wavevectors must be zero to avoid a spontaneous spin current.
A 4-fold lattice symmetry (as in Sr$_3$Ru$_2$O$_7$) suggests four preferred wavevectors. There are two ways to superpose these: in pairs of $\pm {\bf q}$ leading to a spin density wave in one of two directions that breaks the 4-fold rotational symmetry to 2; a superposition of all four symmetry related wavevectors leading to a spin crystal which preserves the lattice symmetry. An example of the latter case is shown in the inset to Fig.~\ref{fig:PhaseDiagram}.

\begin{figure}[bt]
\centerline{\includegraphics[width=2.75in]{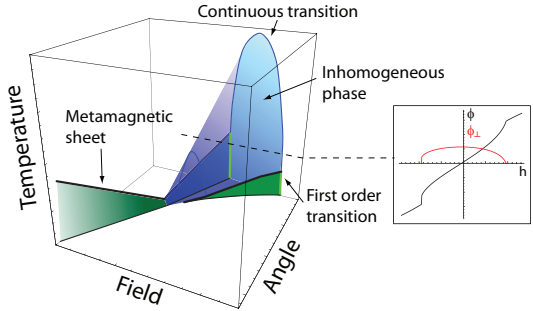}}
\caption{
The phase
  diagram rotated into the experimental orientation. The dashed line shows 
  a trajectory through the inhomogeneous region. The inset shows the variation
  of longitudinal and transverse magnetization through this trajectory.}
\label{fig:Cut}
\end{figure}

Comparison with the experimental phase diagram, Fig.~\ref{fig:Experiment}, is
obtained by
expressing  $R$, $H$ and $K_\perp$ as functions of the experimental
parameters $T$, $\theta$ and $h$. These functions are expected to be
analytic (as is confirmed by their detailed microscopic calculation)
and in the usual spirit of the Ginzburg-Landau expansion their leading
dependence near to the critical point is linear~\cite{beyondlinear}.  Here we choose to expand
about the
point along the line of metamagnetic critical end-points where $K_
\perp$
changes sign. Fig.~\ref{fig:Cut} shows the result of such a
correspondence.
The natural parameters of our microscopic theory are field, temperature and band filling.
 An additional mechanism is required to map from filling to angle. One candidate is spin-orbit coupling~\cite{JeanFrancois} (which leads to an angle dependent Zeeman coupling) together with orbital effects of an in-plane field in a bilayer system.  As the anomalous behaviour of Sr$_3$Ru$_2$O$_7$ only appears in the cleanest samples,
its origin must be sensitive to disorder. Our mechanism shows this sensitivity, since disorder smooths out features in the DoS.

Spatially inhomogeneous magnetic structures lead inevitably to enhanced scattering in certain directions.
In order to fully explain the anisotropy, there must be a mechanism for an
in-plane magnetic field to align the magnetic inhomogeneity. Our simple model does not 
contain such a mechanism. We suggest that its origin 
lies in a modification to the dispersion due to in-plane magnetic field, which breaks the 
symmetry between different orientations of the underlying helices. 
In the 
anomalous phase this magnetic 
inhomogeneity leads to enhanced resistivity. 
With a magnetic field in the c-direction, the inhomogeneity does not break the crystal symmetry (at least macroscopically) and resistivity is isotropic.
As the field is rotated into the plane, the magnetic inhomogeneity no longer preserves the lattice symmetry--- either through the formation of an anisotropic spin crystal or by a preponderance of domains of spin density waves of one orientation. This anisotropy is reflected in resistivity.

Spatial modulation of magnetization should show up as Bragg peaks in elastic neutron scattering
in the anomalous region. 
Unfortunately, no such data exist. There are, however, pseudo-elastic data outside of the anomalous region consistent with fluctuations that would freeze into the type of spin-crystals that we predict~\cite{Antiferromagnet}. 

The mechanism of inhomogeneous magnetic phase formation presented here contrasts with two other proposals:
{\it i.} Spin orbit interactions in systems without a centre of inversion symmetry lead to a Dzyalosinskii-Moriya interaction~\cite{DM} that favors the formation of magnetic spirals~\cite{Bak} and possibly magnetic crystals~\cite{Binz06}. We restrict attention
to systems, such as Sr$_3$Ru$_2$O$_7$, that have a centre of inversion symmetry. {\it ii}. Analysis of quantum fluctuation corrections to the theory of itinerant magnets suggests that they can induce metamagnetism and magnetic inhomogeneity~\cite{Belitz}. Whether such effects are important in Sr$_3$Ru$_2$O$_7$ is unclear. We expect that van Hove singularities are characterized by larger energy scales and provide the dominant mechanism. Others have speculated that the anomalous phase in Sr$_3$Ru$_2$O$_7$ may be a nematic metal with a d-wave distortion of the Fermi surface~\cite{Fradkin07}. The topology of the phase diagram resulting from this distortion should be similar to ours if extended in angle. The main distinction is in the spatial modulation that we predict, which could be probed directly by neutron scattering.

In conclusion, it has long been established that the Stoner model with a peak in the
DoS
can lead to metamagnetism. We have shown that a generic band dispersion 
leads to a bifurcation of this metamagnetism by the intervention of a phase of spatially modulated magnetism analogous to the superconducting LOFF state. This behaviour might have been seen already in Sr$_3$Ru$_2$O$_7$. Indeed, 
our analysis is rather general and its results may have broader applicability.
{\it e.g.} NbFe$_2$~\cite{NbFe2} exhibits a peak in resistivity associated with the bifurcation of a metamagnetic transition and finite wavevector magnetic order and ZrZn$_2$~\cite{ZrZn2} may show similar features.

This work was supported by the Royal Society
and the EPSRC. We are grateful to Gil 
Lonzarich and A.P. Mackenzie for insightful discussions.


\begin{thebibliography}{99}

\bibitem{Fulde} 
P. Fulde and R. A. Ferrel,
{\it Phys. Rev.} {\bf 135,} A550 (1964).

\bibitem{LarkinOvchinnikov} 
A. I. Larkin, and Y. N. Ovchinnikov,
{\it Zh. Eksp. Teor. Fiz.} {\bf 47,} 1136 (1964) [{\it Sov. Phys. JETP} {\bf 20}, 762
(1965)].

\bibitem{cold_atoms} 
D. E. Sheehy and L. Radzihovsky,  
{\it Phys. Rev. Lett.} {\bf 96,} 060401 (2006).

\bibitem{Balents} 
L. Balents, C. Varma, 
{\it Phys. Rev. Lett.} {\bf 84,} 1264 (2000).

\bibitem{Loff_review} 
For a recent review see, e.g. R. Casalbuoni and 
 G. Nardulli,
{\it Rev. Mod. Phys.} {\bf 76,} 263 (2004).

\bibitem{LOFF_exp} 
H. A. Radovan {\it et al.},
{\it Nature} {\bf 425,} 51 (2003); 

\bibitem{LOFF_exp_2}
A. Bianchi {\it et al.}, 
{\it Phys. Rev. Lett.} {\bf 91,} 187004 (2003).

\bibitem{Spivak} 
B. Spivak and S. A. Kivelson, 
{\it Phys. Rev. B} {\bf 70}, 155114 (2004); 
S. A. Kivelson, \emph{et al.},
{\it Nature} {\bf 393}, 550 (1998).

\bibitem{Grigera04} 
S. A. Grigera {\em et al.},
{\it Science} {\bf 306,} 1154 (2004).

\bibitem{Green05} 
A. G. Green {\it et al.},
{\it Phys. Rev. Lett.} {\bf 95,} 086402 (2005).

\bibitem{Borzi07} 
R. A. Borzi {\em et al.},
{\it Science} {\bf 315,} 214 (2007).

\bibitem{Grigera03a} 
S. A. Grigera {\em et al.},
{\it Phys. Rev. B} {\bf 67,} 214427 (2003).

\bibitem{Grigera01} 
S. A. Grigera {\em et al.},
{\it Science} {\bf 294,} 329 (2001).

\bibitem{Wohlfarth} 
E. P. Wohlfarth and P. Rhodes, 
{\it Phil. Mag.} {\bf 7,} 1817 (1962).

\bibitem{Felix} A. Tamai {\it et al.}, 
{\it Phys. Rev. Lett.} {\bf 101}, 026407 (2008).

\bibitem{Binz04} 
B. Binz and M. Sigrist,
{\it Europhys. Lett.} {\bf 65,} 816 (2004),
M. Shimizu,
{\it J. Physique} {\bf 43,} 155 (1982).

\bibitem{Gil} 
P. Monthoux and G. G. Lonzarich,
{\it Phys. Rev. B} {\bf 71,} 054504 (2005).

\bibitem{Rice} 
T. M. Rice,
{\it Phys. Rev. B} {\bf 2,} 3619 (1970).

\bibitem{foot}
$u|{\bf k}+ {\bf q}/2, \uparrow \rangle+ v|{\bf k}-{\bf q}/2
,\downarrow \rangle$ with $u^2+v^2=1$ transform to real space
states with Euler angles $ \theta=\cos^{-1} u$, $\phi={\bf q}\cdot{\bf r}$.

\bibitem{Millis}  
A. J. Millis {\em et al.}
{\it Phys. Rev. Lett.} {\bf 88,} 217204 (2002).

\bibitem{Moriya} 
T. Moriya, ``Spin fluctuations in itinerant electron magnetism'', Springer-Verlag (1985).

\bibitem{foot2}
In the Stoner model,
$
\bar{M}=2h/g+\sum_{{\bf k},\sigma=\pm 1}
\sigma n_{\rm F}(\epsilon_{\bf k}-g\bar{M}\sigma/2),
$
where 
$n_{\rm F}(\epsilon)$ denotes the Fermi  distribution. 

\bibitem{Andrew} A.M. Berridge {\em et al.}
 in preparation.

\bibitem{extra}
In the 2D square lattice, the isotropic contribution to the Ginzburg-Landau free energy density,
$(\partial^2{\bf \phi}_{\perp})^2$, is augmented by a term proportional 
to $(\partial_x^2 {\bf \phi}_{\perp})\cdot(\partial_y^2 {\bf \phi}_{\perp})$.  
  
\bibitem{Kperp}
$K_{\perp}= -\frac{1}{4g\bar M^3 V} \sum_{{\bf k}, \sigma} \left[
\sigma n_F(\epsilon_{{\bf k},\sigma})
+g \bar M  n'_F(\epsilon_{{\bf k},\sigma})/2 
\right]
(\partial_{\bf k} \epsilon_{\bf k})^2$, 
where
$n'_F\equiv \partial_\epsilon n_F$ and $\epsilon_{{\bf k},\sigma}=
\epsilon_{\bf k} - g \bar M \sigma /2$.

\bibitem{beyondlinear}
Whilst our microscopic theory can be used to determine the dependence upon external parameters beyond linear order, the electron dispersion is not known with sufficient accuracy for this to be useful here. We will discuss these non-linear dependencies in a forthcoming longer paper.

\bibitem{JeanFrancois} Jean-Fran\c cois Mercure, PhD thesis (2008)

\bibitem{Antiferromagnet}
S. Hayden and S. Ramos, private communication. See also L. Capogna {\it et al} {\it App. Phys. A} {\bf 74,} 926 (2002), and K. Kitagawa {\it et al}, {\it Phys. Rev. B} {\bf 75}, 024421 (2007).

\bibitem{DM} 
I. Dzyaloshinskii,
{\it J. Phys. Chem. Sol.} {\bf 4,} 241 (1958),
T. Moriya,
{\it Phys. Rev.} {\bf 120,} 91 (1960).

\bibitem{Bak} 
P. Bak and M. H. Jensen,
{\it J. Phys. C: Solid State Phys.} {\bf 13,} L881 (1980),
O. Nakanishi \emph{et al.}
{\it Sol. Stat. Comm.} {\bf 35,} 995 (1980).

\bibitem{Binz06} 
B. Binz, \emph{et al.},
{\it Phys. Rev. Lett.} {\bf 96,} 207202 (2006),
C. Pfleiderer {\it et al.}
{\it Nature} {\bf 427,} 227 (2004),
U. K. R\"ossler, \emph{et al.}, 
{\it Nature} {\bf 442,} 79 (2006).

\bibitem{Belitz} 
D. Belitz, \emph{et al.}, 
{\it Phys. Rev. Lett.} {\bf 82,} 4707 (1999),
D. Belitz, \emph{et al.}, 
{\it Phys. Rev. B} {\bf 73,} 054431 (2006),
J. Rech, \emph{et al.}, 
{\it Phys. Rev. B} {\bf 74,} 195126 (2006).
  
\bibitem{Fradkin07} 
E. Fradkin, \emph{et al.}, 
{\it Science} {\bf 315,} 196 (2007),
H.-Y. Kee  and Y. B. Kim, 
{\it Phys. Rev. B} {\bf 71,} 184402 (2005).

\bibitem{NbFe2} 
M. Brando {\it et al.},
{\it J. Magn. Matter.} {\bf 310,} 852 (2007).

\bibitem{ZrZn2} 
M. Uhlarz, \emph{et al.}, 
{\it Phys. Rev. Lett.} {\bf 93,} 256404 (2004).

\end{thebibliography}
\end{document}